\documentclass{elsart}

\usepackage{graphicx}
\usepackage{amssymb}

\begin{document}
\begin{frontmatter}
\title{Generalization of Euler's summation formula to path integrals\thanksref{MNTR}}
\thanks[MNTR]{Supported by the Ministry of
Science and Environmental Protection of the Republic of
Serbia through projects No. 1486 and 1899.}

\author{A. Bogojevi\' c\corauthref{alex}},
\ead{alex@phy.bg.ac.yu}
\corauth[alex]{Corresponding author.}
\author{A. Bala\v z}, \and
\author{A. Beli\' c}
\address{Scientific Computing Laboratory, Institute of Physics\\
P. O. Box 57, 11001 Belgrade, Serbia and Montenegro}

\begin{abstract}
A recently developed analytical method for systematic improvement of the convergence of path integrals is used to derive a generalization of Euler's summation formula for path integrals. The first $p$ terms in this formula improve convergence of path integrals to the continuum limit from $1/N$ to $1/N^p$, where $N$ is the coarseness of the discretization. Monte Carlo simulations performed on several different models show that the analytically derived speedup holds.
\end{abstract}

\begin{keyword}
Path integral \sep Quantum theory \sep Effective action
\PACS 05.30.-d \sep 05.10.Ln \sep 03.65.Db
\end{keyword}
\end{frontmatter}

\section{Introduction}
\label{sec:intro}

Feynman's path integrals \cite{feynmanhibbs,feynman} represent a compact and rich formalism for dealing with quantum theories. They have proved to be powerful tools for investigating symmetries, deriving non-perturbative results, delineating connections between different theories and different sectors of theories. Their flexibility and intuitive appeal have allowed us to generalize quantization to ever more complicated systems and have led to a rich cross fertilization of ideas between high energy and condensed matter physics \cite{itzyksondrouffe,parisi}. Today, they are used both analytically and numerically \cite{barkerhenderson,barker,kaloswhitlock,pollockceperley,ceperley} in many other areas of physics, in chemistry and materials science, as well as in quantitative finance \cite{kleinert}. 

Unfortunately we still have very little knowledge of the precise mathematical properties of path integrals. As a result only an extremely limited number of path integrals can be solved exactly. Although the functional formalism has been instrumental for deriving many general approximation techniques, along with a host of model-specific approximations, there remain many models of interest that need to be treated numerically (e.g. using Monte Carlo simulations).

Numerical integrations of path integrals have, however, proven to be notoriously demanding of computing time.  
For this reason several research groups have worked at improving the convergence of path integrals. Until recently the best available result for partition functions of a generic $N$-fold discretized theory led to a $1/N^4$ convergence \cite{takahashiimada,libroughton,jangetal}. A new systematic analysis of the relation of discretizations of different coarseness \cite{prl05,prb05} has made it possible to dramatically improve convergence for general transition amplitudes, not just partition functions. A result of this investigation has been a procedure for constructing a series of effective actions $S^{(p)}$ having the same continuum limit as the starting action $S$, but which approach that limit as $1/N^p$. Explicit expressions for these effective actions have so far been constructed for $p\le 9$ \cite{prb05,scl}. In the current paper we will build on this derivation (and simplify it to some extent) and will cast the new analytical input in the form of a generalized Euler summation formula for path integrals. It is our belief that the existence of such a general formula strongly hints at the possibility of (currently unseen) simplifications that might make it possible to set up a rigorous theory of path integration. In section \ref{sec:ordinary} we present a simple derivation of Euler's summation formula for ordinary integrals as a useful guide to the generalization to path integrals that is given in sections \ref{sec:general} and \ref{sec:euler}.

\section{Euler summation formula for ordinary integrals}
\label{sec:ordinary}

The current status of the development of the path integral formalism is quite similar to that of ordinary integrals before the setting up of integration theory by Riemmann. In those days integrals were calculated directly from the defining formula, i.e. one looked at a specific discretization of the integral (Darboux sum), attempted to do the sum explicitly, and finally tried to calculate the continuum limit. For example,
\begin{equation}
I[f]\equiv \int_0^T f(t)dt = \lim_{N\to\infty}I_N[f]\ ,\,\mbox{where}\quad
I_N[f]=\sum_{n=1}^N f(t_n)\epsilon_N\ ,
\end{equation}
$\epsilon_N=T/N$ and $t_n=n\epsilon_N$. 
It goes without saying that done this way, even the simplest ordinary integrals presented a challenge. The mathematicians of the 18th century did not have computers at their disposal or the development of integration theory might have come much later, i.e. they might have succumbed to doing brute force numerical calculations of integrals of all but the simplest functions. The problem with these hypothetical numerical calculations would have been two fold: they would have been inefficient (the discretized sums converge slowly to the continuum value), and they would have worked (thus quite probably slowing down the further development of integration theory). Luckily, this early numerical road was not open. The last great step in the development of integration before Riemmann was made by
Euler.

Discretization is not unique. This makes it possible to change $f(t)$ to some other function (adding terms proportional to $\epsilon_N$, $\epsilon_N^2$, etc.) without changing the integral. Let us assume that $f^*(t)$ is such an equivalent function with the added property that the sums $I_N[f^*]$ do not depend on $N$. In fact we shall present a way of explicitly constructing $f^*(t)$ for any given $f(t)$. 
We first look at the simple case of $f(t)=1$. Now 
\begin{equation}
I_N[1]=\sum_{n=1}^N \epsilon_N = T\ ,
\end{equation}
which is already $N$-independent. Hence, in this case, all the additional terms vanish. Note that $f^*(t)$ is completely determined by the original function $f(t)$ (and by $\epsilon_N$), so that the additional terms necessarily depend only on the derivatives $f'$, $f''$, etc. 

The second step is to take $f(t)=t$. In this case we get
\begin{equation}
I_N[t]=\sum_{n=1}^N t_n\epsilon_N=\frac{N(N+1)}{2}\,\frac{T^2}{N^2}=\frac{T^2}{2}+\frac{T^2}{2N}\ .
\end{equation}
From this it follows that $I_N[t-\frac{\epsilon_N}{2}]=\frac{T^2}{2}$. Therefore, up to $f''$ and higher derivatives of $f$ that all vanish for linear $f(t)$, we have $f^*(t)=f(t) - \frac{\epsilon_N}{2} f'(t)$. 

We continue this procedure by looking at $f(t)=t^2$. In this case we find
\begin{equation}
I_N[t^2]=\sum_{n=1}^N t_n^2\epsilon_N=\frac{N(N+1)(2N+1)}{6}\,\frac{T^3}{N^3}=\frac{T^3}{3}+\frac{T^3}{2N}+\frac{T^3}{6N^2}\ .
\end{equation}
It follows that $I_N[t^2-\epsilon_N t_n-\frac{2}{3} \epsilon_N^2]=\frac{T^3}{3}$. In terms of $f^*$ this gives 
$f^*(t)=f(t) - \frac{\epsilon_N}{2} f'(t) - \frac{2\epsilon_N^2}{3} f''(t)+\ldots$. The additional terms now depend on higher powers of $\epsilon_N$ as well as on higher derivatives and are determined by considering $I_N[t^3]$, and so on. In this way we have constructed a procedure for finding $f^*(t)$ for any given $f(t)$. Remembering that $I_N[f^*]$ does not depend on $N$ we find
\begin{equation}
\int_0^T f(t) dt =\sum_{n=1}^Nf(t_n)\epsilon_N-\frac{\epsilon_N}{2}\sum_{n=1}^N f'(t_n)\epsilon_N - \frac{2\epsilon_N^2}{3}\sum_{n=1}^N f''(t_n)\epsilon_N+\ldots\ .
\end{equation}
This is the well-known Euler summation formula. 
We may also write it more compactly as
\begin{equation}
\label{euler}
I[f]=I_N[f^{(p)}]+O(\epsilon_N^p)\ ,
\end{equation}
where $f^{(p)}$ is the truncation of $f^*$ to the first $p$ terms.
The Euler formula gives the analytical relation between integrals and their discretized sums. Looked at numerically, this formula allows us to increase the speed of convergence of discretized expressions to the continuum limit. For example, in the defining relation the discretized expressions differ from the continuum by a term of order $O(1/N)$. By using the Euler sum formula with $p$ terms we can reduce that error to $O(1/N^p)$. All that is needed to do this is that the integrand is differentiable $p-1$ times. the following sections we will generalize the above approach to path integrals. 

\section{General properties of path integrals}
\label{sec:general}

In the functional formalism the quantum mechanical amplitude $A(a,b;T)=\langle b|e^{-T\hat H}|a\rangle$ is given 
in terms of a path integral which is simply the $N\to\infty$ limit of the $(N-1)$-- fold integral expression
\begin{equation}
\label{amplitudeN}
A_N(a,b;T)=\left(\frac{1}{2\pi\epsilon_N}\right)^{\frac{N}{2}}\int dq_1\cdots dq_{N-1}\,e^{-S_N}\ .
\end{equation}
The Euclidean time interval $[0,T]$ has been subdivided into $N$ equal time steps of length $\epsilon_N=T/N$, with $q_0=a$ and $q_N=b$. $S_N$ is the naively discretized action of the theory. We focus on actions of the form 
\begin{equation}
\label{action}
S=\int_0^Tdt\,\left(\frac{1}{2}\, \dot q^2+V(q)\right)\ ,
\end{equation}
whose naive discretization is simply
\begin{equation}
\label{actionN}
S_N=\sum_{n=0}^{N-1}\left(\frac{\delta_n^2}{2\epsilon_N}+\epsilon_NV_n\right)\ ,
\end{equation}
where $\delta_n=q_{n+1}-q_n$, $V_n=V(\bar q_n)$, and $\bar q_n=\frac{1}{2}(q_{n+1}+q_n)$. We use units in which $\hbar$ and particle mass equal 1.

As was the case with ordinary integrals the definition of the path integrals also makes it necessary to make the transition from the continuum to the discretized theory, a process that is far from unique. For theories described by eq.~(\ref{action}) we have the freedom to choose any point in $[q_n,q_{n+1}]$ in which to evaluate the potential without changing physics -- the discretized amplitudes do differ, but they tend to the same continuum limit. The calculations we present turn out to be simplest in the mid-point prescription where the potential $V$ is evaluated at $\bar q_n$. A more important freedom related to our choice of discretized action has to do with the possibility of introducing additional terms that explicitly vanish in the continuum limit. Actions with such additional terms will be called effective. For example, the term 
$\sum_{n=0}^{N-1}\epsilon_N\,\delta_n^2\,g(\bar q_n)$, 
where $g$ is regular when $\epsilon_N\to0$, does not change the continuum physics since it goes over into $\epsilon_N^2\int^T_0dt\,\dot q^2\,g(q)$, i.e. it vanishes as $\epsilon_N^2$. Such terms do not change the physics, but they do affect the speed of convergence. A systematic study of the relation between different discretizations of the same path integral will allow us to explicitly construct a series of effective actions with progressively faster convergence to the continuum. Before we do this we will parallel the derivation in the previous section and derive some general properties of the best effective action.

The amplitude $A(a,b;T)$ of some theory with action $S$ satisfies
\begin{equation}
\label{linearity}
A(a,b;T)=\int dq_1 \cdots dq_{n-1} A(b,q_{n-1};\epsilon_N)\cdots A(q_1,a;\epsilon_N)\ ,
\end{equation}
for all $N$. This general relation is a direct consequence of the linearity of states in a quantum theory. In analogy with ordinary integrals let us now suppose that there exists an effective action $S^*$ that is equivalent to $S$ (i.e that leads to the same continuum limit for all path integrals) with the additional property that its $N$-fold discretized amplitude $A_N^*(a,b;T)$ does not depend on $N$, i.e. that satisfies
\begin{equation}
\label{stardef}
A_N^*(a,b;T)=A(a,b;T)\ .
\end{equation}
As was the case in the previous section we will in fact construct a general procedure for evaluating this effective action. For actions of the form given in eq. (\ref{action}) we may write the amplitude as
\begin{equation}
A(q_{n+1},q_n;\epsilon_N)=\left(\frac{1}{2\pi\epsilon_N}\right)^\frac{1}{2}\exp \left({-\frac{\delta_n^2}{2\epsilon_N}}\right) {\mathcal A} (q_{n+1},q_n;\epsilon_N)\ ,
\end{equation}
where the reduced amplitude ${\mathcal A}\to 1$ as $\epsilon_N\to 0$. Writing $S_N^*$ as
\begin{equation}
S_N^*=\sum_{n=o}^{N-1}\left(\frac{\delta_n^2}{2\epsilon_N}+\epsilon_N W_n^*\right)\ ,
\end{equation}
and using eq. (\ref{amplitudeN}), (\ref{linearity}) and (\ref{stardef}) we find
\begin{equation}
\exp\left({-\epsilon_N W_n^*}\right)={\mathcal A}(q_{n+1},q_n;\epsilon_N)\ .
\end{equation}
Note that $W_n^*$ is reminiscent of some effective potential, so it should depend on $\bar q_n$, however, from the above relation we see that it must also depend on $\delta_n$. In addition, $W^*$ also has an explicit dependence on the discrete time step $\epsilon_N$, hence 
\begin{equation}
W_n^*=W^*(\delta_n,\bar q_n; \epsilon_N)\ .
\end{equation}
As we have seen, the above functional form is a direct consequence of the linearity of quantum theory. The equivalence of $S$ and $S^*$ implies that $W^*\to V(\bar q)$ when $\epsilon_N$ and $\delta$ go to zero. The final general property of $W^*$ follows from the reality of amplitudes in the Euclidean formalism. Using the hermiticity of the Hamiltonian we find $A(a,b;T)=A(a,b;T)^\dag=\langle b|e^{-T\hat H}|a\rangle ^\dag=  \langle a|e^{-T\hat H}|b\rangle = A(b,a;T)$. In terms of $W^*$ this gives us
\begin{equation}
W^*(\delta_n,\bar q_n;\epsilon_N)=W^*(-\delta_n,\bar q_n;\epsilon_N)\ ,
\end{equation}
or, said another way, only even powers of $\delta_n$ are present in the expansion of $W^*$:
\begin{equation}
W^*(\delta_n,\bar q_n;\epsilon_N)=g_0(\bar q_n;\epsilon_N)+\delta_n^2\,g_1(\bar q_n;\epsilon_N)+\delta_n^4\,g_2(\bar q_n;\epsilon_N)+\ldots\ .
\end{equation}
All the functions $g_k$ are regular in the $\epsilon\to 0$ limit. The link to the starting theory is now simply $g_0(\bar q_n;\epsilon_N)\to V(\bar q_n)$ as $\epsilon_N$ goes to zero. This concludes the general properties of $W^*$. The remaining properties will be analyzed in the following section by studying the relation of discretizations of different coarseness.

\section{Euler summation formula for path integrals}
\label{sec:euler}

We start by studying the relation between the $2N$-fold and $N$-fold discretizations of the same theory. From eq.~(\ref{amplitudeN}) we see that we can write the $2N$-fold amplitude as an $N$-fold amplitude given in terms of a new action $\widetilde S_N$ determined by
\begin{equation}
\label{tildeS}
e^{-\widetilde S_N}=\left(\frac{2}{\pi\epsilon_N}\right)^{\frac{N}{2}}\int dx_1\cdots dx_N\;e^{- S_{2N}}\ ,
\end{equation}
where $S_{2N}$ is the $2N$-fold discretization of the starting action. We have written the $2N$-fold discretized coordinates $Q_0,Q_1,\ldots,Q_{2N}$ in terms of $q$'s and $x$'s in the following way: $Q_{2k}=q_k$ and $Q_{2k-1}=x_k$. Note that we have $q_0=a$, $q_N=b$, while the $N-1$ remaining $q$'s play the role of the dynamical coordinates in the $N$-fold discretized theory. The $x$'s are the $N$ remaining intermediate points that we integrate over in eq.~(\ref{tildeS}). It is not difficult to see that if we use the naively discretized action $S_N$ one obtains for $\widetilde S_N$ an expression that is not of the same form as $S_N$. 

Having in mind the results of the previous section it is best to use the effective action 
\begin{equation}
S_N^*=\sum_{n=o}^{N-1}\left(\frac{\delta_n^2}{2\epsilon_N}+\epsilon_N W^*(\delta_n,\bar q_n; \epsilon_N)\right)\ ,
\end{equation}
which gives the same result for both the $2N$-fold and $N$-fold discretizations. Therefore, in this case we get
\begin{equation}
e^{-S^*_N}=\left(\frac{2}{\pi\epsilon_N}\right)^{\frac{N}{2}}\int dx_1\cdots dx_N\;e^{- S^*_{2N}}\ .
\end{equation}
From this one easily finds
\begin{equation}
\label{integral}
e^{-\epsilon_N W^*(\delta_n,\bar q_n; \epsilon_N)}=
\left(\frac{2}{\pi\epsilon_N}\right)^{\frac{1}{2}}
\int_{-\infty}^{+\infty} dy\;\exp\left(-\frac{2}{\epsilon_N}y^2\right)\;F\left(\bar q_n+y;\frac{\epsilon_N}{2}\right)\ ,
\end{equation}
where
\begin{eqnarray}
\label{F}
\lefteqn{
-\,\frac{2}{\epsilon_N}\,\ln F(x;\epsilon_N)=g_0\Big(\frac{q_{n+1}+x}{2}\,;\epsilon_N\Big)+
g_0\Big(\frac{x+q_n}{2}\,;\epsilon_N\Big)+}\nonumber\\
&&+(q_{n+1}-x)^2\,g_1\Big(\frac{q_{n+1}+x}{2}\,;\epsilon_N\Big)+
(x-q_n)^2\,g_1\Big(\frac{x+q_n}{2}\,;\epsilon_N\Big)+\ldots
\end{eqnarray}

The above integral equation can be solved for the simple cases of a free particle and a harmonic oscillator, and gives the well known results. 
Note however that for a general case the integral in eq.~(\ref{integral}) is in a form that is ideal for an asymptotic expansion \cite{erdelyi}. The time step $\epsilon_N$ is playing the role of small parameter (in complete parallel to the role $\hbar$ plays in standard semi-classical, or loop, expansion). As is usual, the above asymptotic expansion is carried through by first Taylor expanding $F\left(\bar q_n + y;\frac{\epsilon_N}{2}\right)$ around $\bar q_n$ and then by doing the remaining Gaussian integrals. Assuming that $\epsilon_N<1$ (i.e. $N>T$) we have
\begin{eqnarray}
\label{formula}
\lefteqn{g_0(\bar q_n;\epsilon_N)+\delta_n^2\,g_1(\bar q_n;\epsilon_N)+\delta_n^4\,g_2(\bar q_n;\epsilon_N)+\ldots=}\nonumber\\
&&=-\frac{1}{\epsilon_N}\ln\left[\sum_{m=0}^{\infty}\frac{F^{(2m)}\left(\bar q_n;\frac{\epsilon_N}{2}\right)}{(2m)!!}\,\left(\frac{\epsilon_N}{4}\right)^m\right]\ .
\end{eqnarray}
Note that $F^{(2m)}(x;\epsilon_N)$ denotes the corresponding derivative with respect to $x$. All that remains is to calculate these expressions using eq.~(\ref{F}) and to expand all the $g_k$'s around the mid-point $\bar q_n$. This is a straight forward though tedious calculation. In this paper we will illustrate the general procedure for calculating $S^*$ by explicitly giving its expansion to order $\epsilon_N^3$:
\begin{eqnarray}
\label{halvingrelations}
&&g_0(\bar q_n\,;\epsilon_N) =
g_0\left(\bar q_n;\frac{\epsilon_N}{2}\right)+\epsilon_N\left[{\frac{1}{4}}g_1\left(\bar q_n;\frac{\epsilon_N}{2}\right)+\frac{1}{32}g_0''\left(\bar q_n;\frac{\epsilon_N}{2}\right)\right]+\nonumber\\
&&\qquad+\epsilon_N^2\,\left[\frac{3}{16}g_2\left(\bar q_n;\frac{\epsilon_N}{2}\right)-\frac{1}{32}g_0'\,^2\left(\bar q_n;\frac{\epsilon_N}{2}\right)+\frac{1}{2048}g_0^{(4)}\left(\bar q_n;\frac{\epsilon_N}{2}\right)+\right.\nonumber\\
&&\qquad\left.+\frac{3}{128}g_1''\left(\bar q_n;\frac{\epsilon_N}{2}\right)\right]\nonumber\\
&&g_1(\bar q_n\,;\epsilon_N) = \frac{1}{4}g_1\left(\bar q_n;\frac{\epsilon_N}{2}\right)+\frac{1}{32}g_0''\left(\bar q_n;\frac{\epsilon_N}{2}\right)+\\
&&\qquad+\epsilon_N\left[\frac{3}{8}g_2\left(\bar q_n;\frac{\epsilon_N}{2}\right)+\frac{1}{1024}g_0^{(4)}\left(\bar q_n;\frac{\epsilon_N}{2}\right)-\frac{1}{64}g_1''\left(\bar q_n;\frac{\epsilon_N}{2}\right)\right]\nonumber\\
&&g_2(\bar q_n\,;\epsilon_N) = \frac{1}{16}g_2\left(\bar q_n;\frac{\epsilon_N}{2}\right)+\frac{1}{6144}g_0^{(4)}\left(\bar q_n;\frac{\epsilon_N}{2}\right)+\frac{1}{128}g_1''\left(\bar q_n;\frac{\epsilon_N}{2}\right)\ .\nonumber
\end{eqnarray}

In the above relations we expanded $g_0$ up to $\epsilon_N^2$, $g_1$ up to $\epsilon_N$, etc. We also disregarded all the higher $g_k$'s. The reason for this is that the short time propagation of any theory satisfies $\delta_n^2\propto\epsilon_N$ while the $g_k$ term enters the action multiplied by $\delta_n^{2k}$. In general, if we wish to expand the effective action to $\epsilon_N^p$ we need to evaluate only $g_0$ (up to $\epsilon_N^{p-1}$) and the remaining $p-1$ functions $g_k$ (up to $\epsilon_N^{p-1-k}$). The task of calculating the effective action to large powers of $\epsilon_N$ is time-consuming and is best done with the help of a standard package for algebraic calculations such as Mathematica. Using Mathematica we determined the corresponding expressions for $p\le 9$. 

Although the above system of recursive relations is non-linear, it is in fact quite easy to solve if we remember that the system itself was derived via an expansion in $\epsilon_N$. Having this in mind we first write all the functions as expansions in powers of $\epsilon_N$ that are appropriate to the level $p$ we are working at. For $p=3$, we have   
\begin{eqnarray}
\label{expand}
g_0(\bar q_n;\epsilon_N) &=& V(\bar q_n) + \epsilon_N R_1(\bar q_n)+\epsilon_N^2 R_2(\bar q_n)\nonumber\\
g_1(\bar q_n;\epsilon_N) &=& R_3(\bar q_n) + \epsilon_N R_4(\bar q_n)\\
g_2(\bar q_n;\epsilon_N) &=& R_5(\bar q_n)\ .\nonumber
\end{eqnarray}
Putting this into the Eq. (\ref{halvingrelations}) we determine the functions $R_1$ to $R_5$ in terms of $V$. The $p=3$ level solution equals
\begin{eqnarray}
\label{p3continuum}
g_0 &=& V + \epsilon_N\frac{V''}{12} + \epsilon_N^2\left[-\frac{V'\,^2}{24}+\frac{V^{(4)}}{240}\right]\nonumber\\
g_1 &=& \frac{V''}{24} + \epsilon_N\frac{V^{(4)}}{480}\\
g_2 &=& \frac{V^{(4)}}{1920}\ .\nonumber
\end{eqnarray}

Note that $W^*$ depends only on the initial potential $V$ and its derivatives (as well as on $\epsilon_N$). One can similarly calculate the effective action $S^*$ to any desired level $p$. We denote the $p$ level truncation of the effective action as $S^{(p)}$. $S^{(p)}$ has the property that its $N$-fold amplitudes deviate from the continuum expressions as $O(\epsilon_N^p)$ 
\begin{equation}
\label{convergence}
A(a,b;T)=A^{(p)}_N(a,b;T)+O(\epsilon_N^p)\ .
\end{equation}
Comparing this to eq. (\ref{euler}) we see that we have just derived the generalization of the Euler summation formula to path integrals.
Just as with the ordinary Euler formula it gives the relation between path integrals and their discretizations to any given precision.

It is important to note that one solves for the effective action at level $p$ but once for all theories, i.e. the solution that is found holds for all initial potentials. The only requirement for the level $p$ solution is that the starting potential is differentiable $2p-2$ times. Solutions for larger values of $p$ are a bit more cumbersome, however, they are just as easy to use in simulations. We have found that the growth in complexity of the effective actions with increasing $p$ has little effect on computation time for $p\le 4$, while simulations with $p=9$ are roughly ten times slower due to this. However, this is an extremely small price to pay for a gain of eight orders of magnitude in the speed of convergence. Expressions for effective actions up to $p=9$ can be found on our web site \cite{scl}. 
\begin{figure}[!ht]
\centering
\includegraphics[width=13.cm]{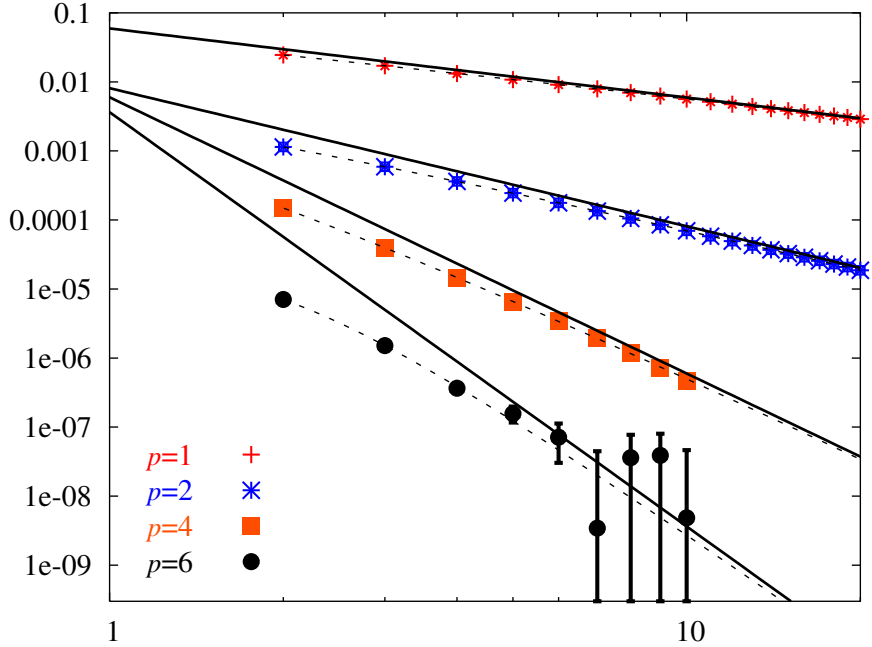}
\caption{\label{p1246} Deviations from the continuum limit $|A_N^{(p)}-A|$ as functions of $N$ for $p=1,2,4$ and $6$ for an anharmonic oscillator with quartic coupling $\lambda=10$, time of propagation $T=1$ from $a=0$ to $b=1$. $N_{MC}$ was $9.2\cdot 10^9$ for $p=1,2$, $9.2\cdot 10^{10}$ for $p=4$, and $3.68\cdot 10^{11}$ for $p=6$. Dashed lines correspond to appropriate $1/N$ polynomial fits to the data. Solid lines give the leading $1/N$ behavior. The level $p$ curve has a $1/N^p$ leading behavior.}
 \end{figure}

The analytical derivations presented work equally well in both the Euclidean and Minkowski formalism (with appropriate $i\epsilon$ regularization), i.e. they are directly applicable to quantum systems as well as to statistical ones. However, the Monte Carlo simulations used to numerically document our analytical results necessarily needed to be done in the Euclidean formalism. We analyzed in detail several models: the anharmonic oscillator with quartic coupling $V=\frac{1}{2}\,q^2+\frac{\lambda}{4!}\, q^4$ and a particle moving in a modified P\"oschl-Teller potential over a wide range of parameters. In all cases we found agreement with eq.~(\ref{convergence}). Fig.~\ref{p1246} illustrates this behavior in the case of an anharmonic oscillator. We see that the $p$ level data indeed differs from the continuum amplitudes as a polynomial starting with $1/N^p$. The deviations from the continuum limit $|A_N^{(p)}-A|$ become exceedingly small for larger values of $p$ making it necessary to use ever larger values of $N_{MC}$ so that the MC statistical error does not mask these extremely small deviations. For $p=6$ we see that although we used an extremely large number of MC samples ($N_{MC}=3.68\cdot 10^{11}$) the statistical errors become of the same order as the deviations already at $N\gtrsim 8$. For $p=9$ this is the case even for $N=2$, i.e. we already get the continuum limit within a MC error of around $10^{-8}$. 

To conclude, we have presented an algorithm that leads to significant speedup of numerical procedures for calculating path integrals. The increase in speed results from new analytical input that comes from a systematic investigation of the relation between discretizations of different coarseness and that leads to a generalization of the Euler summation formula to path integrals. We have presented an explicit procedure for obtaining a set of effective actions $S^{(p)}$ that have the same continuum limit as the starting action $S$, but which approach that limit ever faster. Amplitudes calculated using the $N$-point discretized effective action $S_N^{(p)}$ satisfy $A^{(p)}_N(a,b;T)=A(a,b;T)+O(1/N^p)$. We have obtained and analyzed the effective actions for $p\le 9$ and have documented the speedup up to $1/N^9$ by conducting Monte Carlo simulations of several different models. Several interesting properties 
of this procedure follow from the fact that the solutions were obtained using an asymptotic expansion. These additional properties will be presented and discussed in a future publication. Extension to $d>1$ is also in progress. The derivation of higher dimensional analogues of integral equation ~(\ref{integral}) does not seem to present a problem. The asymptotic expansion used to solve it is also directly generalizable. However, the algebraic recursive relations that determine $S^{(p)}$ will be more complex and may practically limit us to smaller values of $p$.


\begin{thebibliography}{00}

\bibitem{feynmanhibbs}
R. P. Feynman and A. R. Hibbs,
\emph{Quantum Mechanics and Path Integrals}
(McGraw-Hill, New York, 1965).

\bibitem{feynman}
R. P. Feynman,
\emph{Statistical Mechanics}
(W. A. Benjamin, New York, 1972).

\bibitem{itzyksondrouffe}
C. Itzykson and J.-M. Drouffe,
\emph{Statistical Field Theory}
(Cambridge University Press, 1991).

\bibitem{parisi}
G. Parisi,
\emph{Statistical Field Theory}
(Addison Wesley, New York, 1988).

\bibitem{barkerhenderson}
J. A. Barker and D. Henderson,
Rev. Mod. Phys. {\bf 48}, 587 (1976).

\bibitem{barker}
J. A. Barker,
J. Chem. Phys. {\bf 70}, 2914 (1979).

\bibitem{kaloswhitlock}
M. H. Kalos and P. A. Whitlock,
\emph{Monte Carlo Methods}, vol. 1: Basics
(John Wiley and Sons, New York, 1986).

\bibitem{pollockceperley}
E. L. Pollock and D. M. Ceperley,
Phys. Rev. B {\bf 30}, 2555 (1984).

\bibitem{ceperley}
D. M. Ceperley,
Rev. Mod. Phys. {\bf 67}, 279 (1995).

\bibitem{kleinert}
H. Kleinert,
\emph{Path Integrals in Quantum Mechanics, Sta\-tistics, Polymer Physics, and Financial Markets}
(World Scientific, 2004).

\bibitem{takahashiimada}
M. Takahashi and M. Imada,
J. Phys. Soc. Jpn. {\bf 53}, 3765 (1984).

\bibitem{libroughton}
X. P. Li and J. Q. Broughton,
J. Chem. Phys. {\bf 86}, 5094 (1987).

\bibitem{jangetal}
S. Jang, S. Jang, and G. Voth,
J. Chem. Phys. {\bf 115}, 7832 (2001).

\bibitem{prl05}
A. Bogojevi\'c, A. Bala\v z, and A. Beli\'c, 
Phys. Rev. Lett. {\bf 94}, 180403 (2005).

\bibitem{prb05}
A. Bogojevi\'c, A. Bala\v z, and A. Beli\'c, 
Phys. Rev. B {\bf 72}, 064302 (2005).

\bibitem{scl}
http://scl.phy.bg.ac.yu/speedup/

\bibitem{erdelyi}
A. Erdelyi,
\emph{Asymptotic Expansions}
(Dover Publications, New York, 1956).

\end{thebibliography}
\end{document}